\documentclass[sigconf,natbib]{acmart}

\usepackage{multirow}
\usepackage{color, colortbl}
\usepackage{arydshln}
\usepackage{array}

\definecolor{Gray}{gray}{0.9}

\AtBeginDocument{%
  \providecommand\BibTeX{{%
    \normalfont B\kern-0.5em{\scshape i\kern-0.25em b}\kern-0.8em\TeX}}}

\setcopyright{acmcopyright}
\copyrightyear{2023}
\acmYear{2023}
\acmDOI{XXXXXXX.XXXXXXX}

\acmConference[SIGIR '23]{}{July 23--27,
  2023}{Taipei, Taiwan}
\acmPrice{15.00}
\acmISBN{978-1-4503-XXXX-X/18/06}

\begin{document}

\title[Simplifying Content-Based Neural News Recommendation]{Simplifying Content-Based Neural News Recommendation: \protect\\ On User Modeling and Training Objectives}

\author{Andreea Iana}
\affiliation{%
  \institution{University of Mannheim}
  \city{Mannheim}
  \country{Germany}
}
\email{andreea.iana@uni-mannheim.de}

\author{Goran Glavaš}
\affiliation{%
  \institution{CAIDAS, University of Würzburg}
  \city{Würzburg}
  \country{Germany}}
\email{goran.glavas@uni-wuerzburg.de}

\author{Heiko Paulheim}
\affiliation{%
  \institution{University of Mannheim}
  \city{Mannheim}
  \country{Germany}
}
\email{heiko.paulheim@uni-mannheim.de}

\renewcommand{\shortauthors}{Iana, et al.}

\begin{abstract}
The advent of personalized news recommendation 
has given rise to increasingly complex recommender architectures. 
Most neural news recommenders rely on user click behavior and typically introduce dedicated user encoders that aggregate the content of clicked news into user embeddings (\textit{early fusion}). These models are predominantly trained with standard point-wise classification objectives. The existing body of work exhibits two main shortcomings: (1) despite general design homogeneity, direct comparisons between 
models are hindered by varying evaluation datasets and protocols; (2) it leaves alternative model designs and training objectives vastly unexplored.      
In this work, we present a unified framework for 
news recommendation, allowing for a systematic and fair comparison of news recommenders across several crucial design dimensions: (i) 
\textit{candidate-awareness in user modeling}, (ii) \textit{click behavior fusion}, and (iii) \textit{training objectives}. Our findings challenge the status quo in neural news recommendation. We show that replacing sizable user encoders with parameter-efficient dot products between candidate and clicked news embeddings (\textit{late fusion}) often yields substantial performance gains. Moreover, our results render contrastive training a viable alternative to point-wise classification objectives. 
\end{abstract}





\maketitle

\section{Introduction}
\label{sec:intro}

In recent years, content-based news recommendation has seen increasingly complex neural recommender architectures that aim to customize suggestions to users' interests \cite{li2019survey,wu2023personalized}. Most neural news recommendation (\texttt{NNR)} models commonly comprise (i) a dedicated news encoder (NE) and (ii) a user encoder (UE) \cite{wu2021empowering,wu2023personalized}. NEs -- instantiated as a convolutional network \cite{wang2018dkn,wu2019naml,wu2019tanr}, self-attention network \cite{wu2019nrms,wu2020sentirec,qi2021pp}, graph attention network \cite{qi2021kim}, or, most recently, as a pretrained transformer network \cite{wu2021empowering,yu2022tiny} -- convert input features (e.g. titles, categories, entities) into the news embedding. UEs aggregate embeddings of clicked news into a user-level representation by means of sequential \cite{an2019lsturl,qi2020privacy,wang2022mins} or attentive \cite{wu2019naml,wu2019nrms,wang2018dkn} encoders that contextualize embeddings of clicked news based on patterns in clicking behavior \cite{okura2017embedding,an2019lsturl,wu2022temprec}. 
We dub this predominant paradigm \textit{early fusion} (\texttt{EF}) because it aggregates representations of clicked news (i.e., builds user representation) before 
comparison with the recommendation candidate.    

Most \texttt{NNR} models encode users and candidate news separately, in a \textit{candidate-agnostic} manner \cite{an2019lsturl,wu2019naml,wu2019nrms}. \textit{Candidate-aware} models \cite{wang2020fim,qi2021hierec,zhang2021amm,qi2022caum}, in contrast, acknowledge that not all clicked news are equally informative w.r.t. the relevance of the candidate (e.g., a candidate is often representative of only a subset of a user's preferences), and contextualize representations of clicked news with the embedding of the candidate in user-level aggregation with UE. 
Finally, the candidate's embedding (output of NE) is compared against the user embedding (output of UE): the candidate's recommendation score is computed directly as the dot product of the two embeddings \cite{wu2019naml} or with a feed-forward scorer \cite{wang2018dkn}. 
\texttt{NNR} models are predominantly trained via standard classification objectives \cite{wang2018dkn,wu2019naml,wu2019nrms,wu2021empowering} with negative sampling \cite{huang2013learning,wu2019npa}.

The existing body of work has two main shortcomings. First, despite general design homogeneity, direct comparisons between recent \texttt{NNR}s are hindered by lack of transparency and adoption of ad-hoc evaluation protocols \cite{iana2022survey,raza2022news}. In particular, a vast majority of personalized news recommenders are evaluated on proprietary datasets (e.g. MSN News \cite{wu2019naml,wu2019nrms}, Bing News \cite{wang2018dkn}, NewsApp \cite{qi2022caum}). Even the few models evaluated using the publicly available datasets such as Adressa \cite{gulla2017adressa} or MIND \cite{wu2020mind} cannot be directly compared due to different dataset splits and evaluation protocols (e.g., model selection strategy) \cite{zhang2021amm,gong2022positive,wu2021empowering,wang2022mins}. 
Secondly, simpler and arguably more intuitive design alternatives have largely been left unexplored. First, the existing work adopts \texttt{EF} as default architecture, proposing increasingly complex user encoding components \cite{an2019lsturl,qi2022caum}, often with little empirical justification for added complexity.  
Second, only a small fraction of \texttt{NNR}s leverage contrastive learning objectives \cite{ijcai2022infonce,yu2022tiny}, despite such training criteria being proven highly effective in closely related retrieval and recommendation tasks \cite{li2021more,wei2021contrastive,yang2022supervised,xie2022contrastive}.

In this work, we remedy the above shortcomings of current \texttt{NNR}s and shed new light on user modeling and training objectives.\footnote{\textbf{Disclaimer:} In this work we focus exclusively on NNR models that do not resort to graph-based modeling of relations between users.} 
\textbf{1)} Concretely, we introduce a unified framework for neural news recommendation, facilitating systematic and fair comparison of NNR models across three crucial design dimensions: (i) \textit{candidate-awareness in user modeling}, (ii) \textit{click behavior fusion}, and (iii) \textit{training objectives}. \textbf{2)} We propose to replace user modeling with complex user encoders (i.e., \textit{early fusion}) with simple pooling of dot-product scores between candidate and clicked news embeddings (i.e. \textit{late fusion}). We show that, despite conceptual simplicity, \texttt{LF} brings substantial performance gains over \texttt{EF}-based \texttt{NNR}, rendering complex UEs empirically unjustified. \textbf{3)} Finally, we demonstrate the benefits of supervised contrastive training as a viable alternative to point-wise classification. Our work fundamentally challenges the status quo of \texttt{NNR} by introducing simpler and more effective alternatives to the established paradigm based on complex user modeling.

\section{Methodology}
\label{sec:methodology}

\begin{figure}[t]
  \centering
  \includegraphics[width=\linewidth]{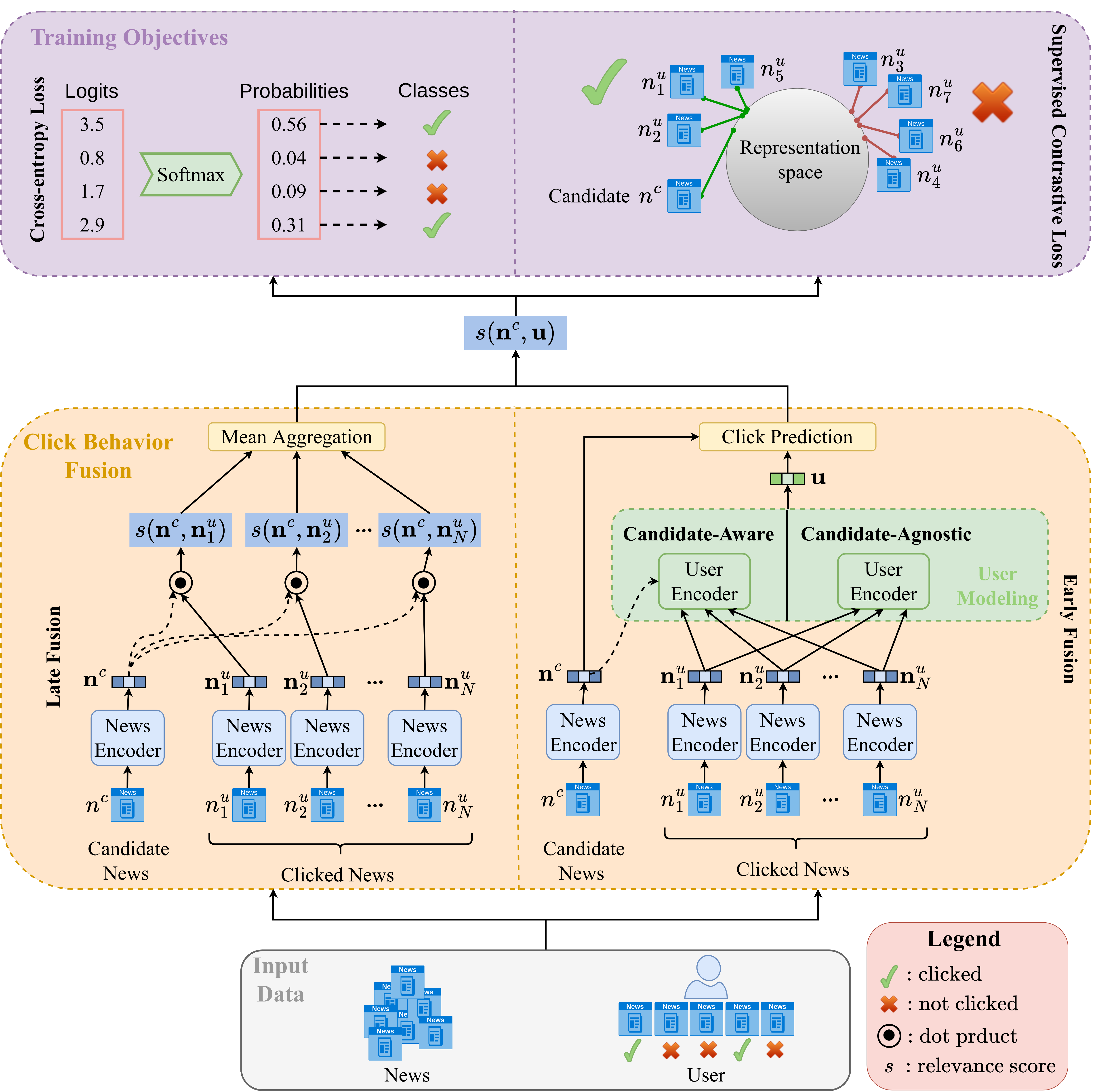}
  \caption{Illustration of the unified \texttt{NNR} framework, focusing on three crucial design dimensions: (i) candidate-awareness in user modeling (green box), (ii) click behavior fusion (orange box), and (iii) training objectives (purple box).}
  \label{fig:framework}
  \vspace{-1em}
\end{figure}

Figure \ref{fig:framework} depicts our unified evaluation framework for \texttt{NNR}, focusing on three critical dimensions of news recommendation. Given input data, comprising news and user behaviors, we analyze (i) candidate-agnostic (\texttt{C-AG}) vs. candidate-aware (\texttt{C-AW}) user modeling under (ii) two click behavior fusion strategies, namely \texttt{EF} and \texttt{LF}, where each model can be (iii) trained by minimizing either the standard cross-entropy loss (\texttt{CE}) or a supervised contrastive objective (\texttt{SCL)}. Next, we describe the models selected for evaluation and formalize the concrete design choices.

\subsection{User Modeling}
\label{subsec:user_modeling_nnr}

\vspace{1.4mm} \noindent \textbf{Candidate-Agnostic (\texttt{C-AG}) Models.} For these models, the UE produces the user embedding from embeddings of clicked news without contextualization against the candidate. We evaluate the following \texttt{C-AG} models, mutually differing in their NE component (i.e., how they embed the clicked news): (1) \textit{NPA} \cite{wu2019npa} uses a personalized attention module to aggregate the representations of the users' clicked news, with projected embeddings of the users IDs as attention queries; (2) \textit{NAML} \cite{wu2019naml} uses additive attention \cite{bahdanau2014neural} to encode users' preferences; (3) \textit{NRMS} \cite{wu2019nrms} learns user representations with a two-layer encoder that consists of  multi-head self-attention \cite{vaswani2017attention} and additive attention; (4) \textit{LSTUR} \cite{wu2020mind} learns user representations with recurrent networks: a short-term user embedding is produced from the clicked news with a GRU \cite{cho2014learning}, and combined with a long-term embedding, consisting of a randomly initialized and fine-tuned part; the final user embedding is then obtained either (i) as the final hidden state of the short-term GRU, initialized with the long-term embedding (\textit{LSTUR\textsubscript{ini}}), or (ii) by simply concatenating the short- and long-term user embeddings (\textit{LSTUR\textsubscript{con}}); (5) \textit{CenNewsRec}  \cite{qi2020privacy} adopts a similar UE architecture as LSTUR, but learns long-term user vectors from clicked news using a sequence of multi-head self-attention and attentive pooling networks, as opposed to storing an explicit embedding per user; (6) \textit{MINS} \cite{wang2022mins} encodes users through a combination of multi-head self-attention, multi-channel GRU-based recurrent network, and additive attention.

\vspace{1.4mm} \noindent \textbf{Candidate-Aware (\texttt{C-AW}) Models.}
UEs in candidate-agnostic models produce the same user embedding, regardless of the content of the candidate news. In contrast, UEs of candidate-aware models, two of which we include in our empirical analysis, produce user embeddings dependent on the candidate. 
(7) \textit{DKN} \cite{wang2018dkn} computes candidate-aware representations of users as the weighted sum of their clicked news embeddings, with weights being produced by an attention network that takes as input the embeddings of the candidate and of the clicked news, as produced by the NE. 
More recently, (8) \textit{CAUM} \cite{qi2022caum} combines (i) a candidate-aware self-attention network to model long-range dependencies between clicked news, conditioned on the candidate, and (ii) a candidate-aware convolutional network (CNN) to capture short-term user interests from adjacent clicks, again conditioned with the candidate's content; the candidate-aware user embedding is finally obtained by attending over the long-range and short-term representations. 

\vspace{1.4mm}
\noindent \textbf{News Encoders.} The \texttt{NNR} models included in our evaluation primarily use news titles as input, which they typically embed via pretrained word embeddings \cite{pennington2014glove}. NAML, LSTUR, MINS, and CAUM additionally leverage category information, with categories embedded with a linear layer. CAUM additionally encodes title entities and DKN exploits knowledge graph embeddings \cite{ji2015knowledge}. The shallow word and entity embeddings are contextualized either using a combination of multi-head self-attention (in NRMS, MINS, CAUM), or a sequence of CNN \cite{kim-2014-convolutional} and additive attention networks (in NAML, LSTUR). NPA \cite{wu2019npa} also utilizes a CNN to contextualize word embeddings, followed by a personalized attention module, analogous to the one used in its user encoder, whereas DKN employs a word-entity-aligned knowledge-aware CNN \cite{wang2018dkn}. CenNewsRec \cite{qi2020privacy} combines the CNN network with multi-head self-attention and additive attention modules. Models with multiple feature vectors produce final news embeddings by simply concatenating them (LSTUR, CAUM), or by attending over them (NAML, MINS).

\subsection{Click Behavior Fusion}
\label{subsec:behavior_fusion}
We question whether the design and computational complexity of \textit{early fusion} (\texttt{EF}), i.e., existence of dedicated user encoders in state-of-the-art \texttt{NNR} models, is justified. To this end, we propose, as a lightweight alternative, the \textit{late fusion} (\texttt{LF}) approach that replaces the elaborate user encoders with mean-pooling of dot-product scores between the embedding of the candidate $n^c$ and the embeddings of the clicked news $n^u_i$. Given a candidate news $n^c$ and a sequence of news clicked by the user $H={n^u_1, ..., n^u_N}$, we compute the relevance score of the candidate news with regards to the user $u$'s history as $s(\mathbf{n}^c, u) = \frac{1}{N} \sum_{i=1}^N \mathbf{n}^c \cdot \mathbf{n}^u_i$, where $\mathbf{n}$ denotes the embedding of a news learned by the news encoder and $N$ the history length. 

Although \texttt{LF} suggests that explicitly encoding user behavior may not be necessary for click prediction, user embeddings are still needed in collaborative-filtering models \cite{li2019survey}. Note that the \texttt{LF} formulation above is equivalent to the dot product between the candidate embedding $\mathbf{n}^c$ and the mean of embeddings of the user's clicked news $\mathbf{n}^u_i$, $s(\mathbf{n}^c, u) = \mathbf{n}^c \cdot \left(\frac{1}{N} \sum_{i=1}^N \mathbf{n}^u_i \right)$. This means that \texttt{LF} can also seamlessly provide user embeddings (simply as averages of clicked news embeddings) if needed
. \texttt{LF} can thus been seen as a \textit{parameterless} user encoder, i.e., a computationally efficient alternative to complex parameterized UEs in existing \texttt{EF} models. Because (i) we produce embeddings of candidate and clicked news independently, and (ii) yield user embeddings as averages of clicked news embeddings, \texttt{LF} models are candidate agnostic (\texttt{C-AG}).

\subsection{Training Objectives}
\label{subsec:training_objectives}

The vast majority of existing \texttt{NNR} work, regardless of the concrete user modeling architecture, tunes the parameters by minimizing the arguably most straightforward classification objective, cross-entropy loss (with negative sampling; see Figure \ref{fig:framework}), and largely fails to explore effective alternatives, foremost contrastive objectives \cite{oord2018representation,ijcai2022infonce}.
This prevents understanding of models effectiveness under different training regimes. We address this limitation by training all models (see \S\ref{subsec:user_modeling_nnr}) not only with (1) common cross-entropy loss (with negative sampling), but also via (2) a contrastive learning objective, in particular supervised contrastive loss \cite{khosla2020supervised}. 

\section{Experimental Setup}
\label{sec:experimental_setup}

\vspace{1.4mm} \noindent \textbf{Data.}
We conduct experiments on the MINDsmall and MINDlarge datasets, introduced by \citet{wu2020mind}. Table \ref{tab:mind_dataset} summarizes their main statistics. Since \citet{wu2020mind} do not release test set labels, we use the respective validation portions for testing, and split the respective training sets into temporally disjoint training (first four days of data) and validation portions (the last day). 
 
\newcolumntype{g}{>{\columncolor{Gray}}r}
\begin{table}
\def\arraystretch{0.9}
  \caption{Statistics of the MINDsmall and MINDlarge datasets.}
  \label{tab:mind_dataset}
  \begin{tabular}{lrg|rg}
    \toprule
     & \multicolumn{2}{c}{\textbf{MINDsmall}} & \multicolumn{2}{c}{\textbf{MINDlarge}} \\ 
     \cmidrule(lr){2-3} \cmidrule(lr){4-5}
    
    \multicolumn{1}{l}{Statistic} & \multicolumn{1}{r}{Train} & \multicolumn{1}{r|}{Test} & \multicolumn{1}{r}{Train} & \multicolumn{1}{r}{Test} \\\hline
    
    \# News & 51,282 & 42,416 & 101,527 & 72,023 \\
    \# Users & 49,108 & 48,593 & 698,365 & 248,973 \\
    \# Impressions & 153,727 & 70,938 & 2,186,683 & 365,201 \\
    \# Categories & 17 & 17 & 18 & 17 \\
    \# Subcategories & 264 & 252 & 285 & 269 \\ 
    
  \bottomrule
\end{tabular}
\end{table}

\vspace{1.4mm} \noindent \textbf{Implementation and Optimization Details.}
We use 300-dimensi\-onal pretrained Glove embeddings \cite{pennington2014glove} and 100-dimensional TransE embeddings \cite{bordes2013translating} pretrained on Wikidata to initialize respectively the word and entity embeddings of the \texttt{NNR} models under comparison. We set the maximum history length to 50. Following \citet{ijcai2022infonce}, our negative sampling creates four negatives per positive example. We find the optimal temperature for \texttt{SCL} using the validation performance, sweeping the interval $[0.08, 0.3]$ with a 0.02 step. We train with batch size of 512 for all \texttt{C-AG} models, 256 for DKN and only 64 for CAUM (due to computational limitations). We set all other model-specific hyperparameters, to optimal values reported in the respective papers.  
We train all models with mixed precision, under a fixed computational budget: for 25 epochs on MINDsmall and 10 epochs on MINDlarge. We optimize with the Adam algorithm \cite{kingma2014adam}, with the learning rate set to 1\textit{e}-4. We repeat each experiment five times (with different random seeds) and report averages (and std. deviation) for common metrics: AUC, MRR, nDCG@5, and nDCG@10.
Each model is trained on a single NVIDIA Tesla V100 GPU with 32GB memory. Our implementation is publicly available.\footnote{Code available at:   \href{https://github.com/andreeaiana/simplifying_nnr}{https://github.com/andreeaiana/simplifying\_nnr}} 

\section{Results and Discussion}
\label{sec:results_discussion}

\begin{table*}[t]
\centering

\caption{Recommendation performance of the compared models under combinations of click behavior fusion (CBF), and training objectives. We report averages and standard deviations across five different runs.}
\label{tab:reults}

\resizebox{\textwidth}{!}{%
    \begin{tabular}{ll|cc:cc:cc:cc|cc:cc:cc:cc}
        \toprule
        \multicolumn{1}{c}{} & \multicolumn{1}{c}{} & \multicolumn{8}{c}{\textbf{MINDsmall}} & \multicolumn{8}{c}{\textbf{MINDlarge}} \\  
        \cmidrule(lr){3-10} \cmidrule(lr){11-18}
        
        \multicolumn{1}{c}{} & \multicolumn{1}{c}{} & \multicolumn{2}{c}{AUC} & \multicolumn{2}{c}{MRR} & \multicolumn{2}{c}{nDCG@5} & \multicolumn{2}{c}{nDCG@10} & \multicolumn{2}{c}{AUC} & \multicolumn{2}{c}{MRR} & \multicolumn{2}{c}{nDCG@5} & \multicolumn{2}{c}{nDCG@10} \\ 
         \cmidrule(lr){3-4} \cmidrule(lr){5-6} \cmidrule(lr){7-8} \cmidrule(lr){9-10}
         \cmidrule(lr){11-12} \cmidrule(lr){13-14} \cmidrule(lr){15-16} \cmidrule(lr){17-18}
         
        \multicolumn{1}{l}{Model} & \multicolumn{1}{l|}{CBF} 
        & \multicolumn{1}{c}{CE} & \multicolumn{1}{c}{SCL} & \multicolumn{1}{c}{CE} & \multicolumn{1}{c}{SCL} & \multicolumn{1}{c}{CE} & \multicolumn{1}{c}{SCL} & \multicolumn{1}{c}{CE} & \multicolumn{1}{c|}{SCL} 
        & \multicolumn{1}{c}{CE} & \multicolumn{1}{c}{SCL} & \multicolumn{1}{c}{CE} & \multicolumn{1}{c}{SCL} & \multicolumn{1}{c}{CE} & \multicolumn{1}{c}{SCL} & \multicolumn{1}{c}{CE} & \multicolumn{1}{c}{SCL} \\ \hline

        & EF 
        & 54.7$\pm$0.6 & 56.5$\pm$0.7 
        & 29.0$\pm$0.7 & 28.4$\pm$0.6 
        & 26.9$\pm$0.8 & 26.6$\pm$0.6 
        & 33.2$\pm$0.8 & 32.6$\pm$0.4 

        & 56.8$\pm$0.2 & 58.1$\pm$0.8 
        & 31.4$\pm$0.5 & 30.0$\pm$0.6 
        & 29.5$\pm$0.4 & 27.7$\pm$0.6 
        & 35.9$\pm$0.4 & 34.3$\pm$0.6 
        \\
        
        \multirow{-2}{*}{NPA} 
       
        & \cellcolor{Gray} LF
        & \cellcolor{Gray} 55.1$\pm$0.9 
        & \cellcolor{Gray} 57.3$\pm$1.2 
        
        & \cellcolor{Gray} 28.6$\pm$0.3 
        & \cellcolor{Gray} 27.5$\pm$1.0 
        
        & \cellcolor{Gray} 26.4$\pm$0.4 
        & \cellcolor{Gray} 25.5$\pm$1.1 
        
        & \cellcolor{Gray} 32.9$\pm$0.4 
        & \cellcolor{Gray} 31.8$\pm$0.9 

        & \cellcolor{Gray} 61.2$\pm$0.5 
        & \cellcolor{Gray} 58.7$\pm$0.8 
        
        & \cellcolor{Gray} 32.1$\pm$0.6 
        & \cellcolor{Gray} 28.3$\pm$0.5 
        
        & \cellcolor{Gray} 30.2$\pm$0.7 
        & \cellcolor{Gray} 26.0$\pm$0.7 
        
        & \cellcolor{Gray} 36.6$\pm$0.6 
        & \cellcolor{Gray} 32.7$\pm$0.7 
        \\ \cdashline{1-18}

        & EF 
        & 50.1$\pm$0.0 & 57.1$\pm$1.1 
        & 33.6$\pm$0.5 & 32.2$\pm$0.7 
        & 31.6$\pm$0.6 & 30.4$\pm$0.7 
        & 38.0$\pm$0.5 & 36.8$\pm$0.7 

        & 50.1$\pm$0.0 & 60.4$\pm$0.6 
        & 33.2$\pm$0.4 & 34.2$\pm$0.4 
        & 31.3$\pm$0.5 & 32.5$\pm$0.3 
        & 37.9$\pm$0.4 & 38.9$\pm$0.3 
        \\
        
        \multirow{-2}{*}{NAML} 
        
        & \cellcolor{Gray} LF 
        & \cellcolor{Gray} 50.0$\pm$0.0
        & \cellcolor{Gray} 62.7$\pm$0.5 
        
        & \cellcolor{Gray} 33.7$\pm$0.8 
        & \cellcolor{Gray} 32.0$\pm$0.7 
        
        & \cellcolor{Gray} 31.8$\pm$0.9 
        & \cellcolor{Gray} 30.3$\pm$0.7 
        
        & \cellcolor{Gray} 38.1$\pm$0.8 
        & \cellcolor{Gray} 36.6$\pm$0.6 

        & \cellcolor{Gray} 50.0$\pm$0.0 
        & \cellcolor{Gray} 65.4$\pm$0.5 
        
        & \cellcolor{Gray} 32.7$\pm$0.5 
        & \cellcolor{Gray} 33.5$\pm$0.5 
        
        & \cellcolor{Gray} 31.0$\pm$0.5 
        & \cellcolor{Gray} 31.7$\pm$0.5 
        
        & \cellcolor{Gray} 37.6$\pm$0.4 
        & \cellcolor{Gray} 38.2$\pm$0.4 
        \\ \cdashline{1-18}

        & EF
        & 52.6$\pm$1.3 & 59.9$\pm$0.6 
        & 27.6$\pm$0.8 & 29.2$\pm$0.7 
        & 25.7$\pm$0.5 & 27.2$\pm$0.9 
        & 32.3$\pm$0.5 & 33.7$\pm$0.7 

        & 54.6$\pm$1.4 & 62.8$\pm$0.7 
        & 31.9$\pm$1.0 & 32.4$\pm$0.5 
        & 30.0$\pm$1.1 & 30.5$\pm$0.7 
        & 36.6$\pm$1.0 & 36.9$\pm$0.6 
        \\
        
        \multirow{-2}{*}{NRMS} 
        
        & \cellcolor{Gray} LF
        & \cellcolor{Gray} 58.9$\pm$1.0 
        & \cellcolor{Gray} 60.2$\pm$1.1 
       
        & \cellcolor{Gray} 31.8$\pm$0.7 
        & \cellcolor{Gray} 30.7$\pm$0.6 
        
        & \cellcolor{Gray} 29.9$\pm$0.7 
        & \cellcolor{Gray} 28.7$\pm$0.6 
        
        & \cellcolor{Gray} 36.3$\pm$0.6 
        & \cellcolor{Gray} 35.1$\pm$0.6 

        & \cellcolor{Gray} 56.1$\pm$2.1 
        & \cellcolor{Gray} 63.6$\pm$1.1 
        
        & \cellcolor{Gray} 32.9$\pm$0.7 
        & \cellcolor{Gray} 32.4$\pm$0.6 
        
        & \cellcolor{Gray} 31.7$\pm$1.1 
        & \cellcolor{Gray} 30.6$\pm$0.8 
        
        & \cellcolor{Gray} 37.8$\pm$0.4 
        & \cellcolor{Gray} 37.1$\pm$0.7 
        \\ \cdashline{1-18}

        & EF\textsubscript{ini}
        & 53.5$\pm$1.2 & 55.4$\pm$0.5 
        & 29.6$\pm$0.5 & 28.1$\pm$0.7 
        & 28.0$\pm$0.5 & 26.5$\pm$0.7 
        & 34.4$\pm$0.4 & 32.9$\pm$0.6 

        & 50.0$\pm$0.1 & 56.9$\pm$1.5 
        & 32.5$\pm$2.4 & 31.3$\pm$1.5 
        & 30.9$\pm$2.4 & 29.8$\pm$1.8 
        & 37.4$\pm$2.4 & 36.2$\pm$1.6 
        \\
        
        & EF\textsubscript{con}
        & 50.2$\pm$0.0 & 59.8$\pm$1.4 
        & 31.8$\pm$0.7 & 31.3$\pm$0.8 
        & 30.1$\pm$0.8 & 30.3$\pm$1.3 
        & 36.4$\pm$0.7 & 36.2$\pm$0.7 

        & 51.4$\pm$0.4 & 54.3$\pm$0.4 
        & 27.7$\pm$0.4 & 26.5$\pm$0.2 
        & 25.9$\pm$0.5 & 24.6$\pm$0.2 
        & 32.3$\pm$0.5 & 31.1$\pm$0.2 
        \\ 
        
        \multirow{-3}{*}{LSTUR}
        
        & \cellcolor{Gray} LF 
        & \cellcolor{Gray} 50.0$\pm$0.0 
        & \cellcolor{Gray} 50.0$\pm$0.0 
        
        & \cellcolor{Gray} 33.8$\pm$0.6 
        & \cellcolor{Gray} 33.9$\pm$0.6 
        
        & \cellcolor{Gray} 31.9$\pm$0.7 
        & \cellcolor{Gray} 32.0$\pm$0.7 
        
        & \cellcolor{Gray} 38.0$\pm$0.6 
        & \cellcolor{Gray} 38.1$\pm$0.6 

        & \cellcolor{Gray} 50.0$\pm$0.0 
        & 50.0$\pm$0.0 
        
        & \cellcolor{Gray} 34.7$\pm$0.6 
        & \cellcolor{Gray} 33.1$\pm$0.2 
        
        & \cellcolor{Gray} 33.2$\pm$0.6 
        & \cellcolor{Gray} 31.6$\pm$0.4 
        
        & \cellcolor{Gray} 39.2$\pm$0.5 
        & \cellcolor{Gray} 37.7$\pm$0.3 
        \\ \cdashline{1-18}

        & EF 
        & 54.0$\pm$0.8 & 60.0$\pm$0.4 
        & 28.3$\pm$0.5 & 30.6$\pm$0.8 
        & 26.5$\pm$0.4 & 28.5$\pm$0.8 
        & 32.9$\pm$0.3 & 34.8$\pm$0.7 

        & 56.4$\pm$0.8 & 64.5$\pm$0.4  
        & 33.7$\pm$0.3 & 33.5$\pm$0.4 
        & 31.9$\pm$0.3 & 31.8$\pm$0.5 
        & 38.3$\pm$0.2 & 38.1$\pm$0.4 
        \\
        
        \multirow{-2}{*}{CenNewsRec} 
        
        & \cellcolor{Gray} LF 
        & \cellcolor{Gray} 59.3$\pm$0.6 
        & \cellcolor{Gray} 61.9$\pm$0.7 
        
        & \cellcolor{Gray} 32.8$\pm$0.8 
        & \cellcolor{Gray} 32.2$\pm$0.8 
        
        & \cellcolor{Gray} 30.9$\pm$0.8 
        & \cellcolor{Gray} 30.4$\pm$0.8 
        
        & \cellcolor{Gray} 37.1$\pm$0.6 
        & \cellcolor{Gray} 36.6$\pm$0.7 

        & \cellcolor{Gray} 53.3$\pm$0.7 
        & \cellcolor{Gray} 64.2$\pm$0.6 
        
        & \cellcolor{Gray} 33.2$\pm$0.4 
        & \cellcolor{Gray} 33.3$\pm$0.4 
        
        & \cellcolor{Gray} 31.4$\pm$0.5 
        & \cellcolor{Gray} 31.7$\pm$0.4 
        
        & \cellcolor{Gray} 37.9$\pm$0.4 
        & \cellcolor{Gray} 38.1$\pm$0.4 
        \\ \cdashline{1-18}

        & EF 
        & 50.6$\pm$0.3 & 62.9$\pm$1.7 
        & 33.7$\pm$1.0 & 32.4$\pm$0.3 
        & 31.9$\pm$1.1 & 30.7$\pm$0.4 
        & 38.3$\pm$0.9 & 37.1$\pm$0.3 

        & 51.7$\pm$0.2 & 65.8$\pm$0.5 
        & 34.3$\pm$0.2 & 34.4$\pm$0.5 
        & 32.5$\pm$0.4 & 32.6$\pm$0.5 
        & 39.1$\pm$0.4 & 39.1$\pm$0.5 
        \\ 
        
        \multirow{-2}{*}{MINS} 
        
        & \cellcolor{Gray} LF 
        & \cellcolor{Gray} 59.1$\pm$1.2 
        & \cellcolor{Gray} 64.2$\pm$0.7 
        
        & \cellcolor{Gray} 35.0$\pm$0.5 
        & \cellcolor{Gray} 34.1$\pm$0.6 
        
        & \cellcolor{Gray} 33.2$\pm$0.6 
        & \cellcolor{Gray} 32.3$\pm$0.7 
        
        & \cellcolor{Gray} 39.4$\pm$0.6 
        & \cellcolor{Gray} 38.5$\pm$0.6 

        & \cellcolor{Gray} 53.8$\pm$0.6 
        & \cellcolor{Gray} 66.7$\pm$0.8 
        
        & \cellcolor{Gray} 34.9$\pm$0.2 
        & \cellcolor{Gray} 34.8$\pm$0.7 
        
        & \cellcolor{Gray} 33.0$\pm$0.2 
        & \cellcolor{Gray} 33.1$\pm$0.7 
        
        & \cellcolor{Gray} 39.5$\pm$0.2 
        & \cellcolor{Gray} 39.6$\pm$0.6 
        \\ \hline 

        & EF 
        & 50.0$\pm$0.0 & 51.0$\pm$2.3 
        & 26.4$\pm$0.6 & 25.9$\pm$0.9 
        & 24.4$\pm$0.7 & 23.9$\pm$1.1 
        & 31.0$\pm$0.6 & 30.5$\pm$0.9 

        & 50.0$\pm$0.0 & 50.0$\pm$0.0 
        & 25.2$\pm$0.4 & 24.8$\pm$0.3 
        & 23.4$\pm$0.7 & 22.6$\pm$0.3 
        & 30.0$\pm$0.5 & 29.1$\pm$0.3 
        \\
        
        \multirow{-2}{*}{DKN} 
        
        & \cellcolor{Gray} LF 
        & \cellcolor{Gray} 50.0$\pm$0.0 
        & \cellcolor{Gray} 50.0$\pm$0.0 
        
        & \cellcolor{Gray} 27.5$\pm$0.6 
        & \cellcolor{Gray} 26.4$\pm$0.8 
        
        & \cellcolor{Gray} 25.0$\pm$0.5 
        & \cellcolor{Gray} 24.0$\pm$0.8 
        
        & \cellcolor{Gray} 31.7$\pm$0.6 
        & \cellcolor{Gray} 30.8$\pm$0.8 

        & \cellcolor{Gray} 50.0$\pm$0.0 
        & \cellcolor{Gray} 50.0$\pm$0.0 
        
        & \cellcolor{Gray} 29.1$\pm$0.4 
        & \cellcolor{Gray} 27.8$\pm$1.1 
        
        & \cellcolor{Gray} 26.3$\pm$0.3 
        & \cellcolor{Gray} 25.4$\pm$1.0 
        
        & \cellcolor{Gray} 33.2$\pm$0.4 
        & \cellcolor{Gray} 32.1$\pm$1.0 
        \\ \cdashline{1-18}

        & EF 
        & 61.4$\pm$1.0 & 63.2$\pm$0.9 
        & 33.8$\pm$0.6 & 33.7$\pm$0.8 
        & 32.0$\pm$0.6 & 31.8$\pm$0.9 
        & 38.4$\pm$0.5 & 38.2$\pm$0.8 

        & 67.1$\pm$0.8 & 66.4$\pm$0.9 
        & 35.3$\pm$0.5 & 35.1$\pm$0.5 
        & 33.6$\pm$0.6 & 33.4$\pm$0.6 
        & 40.1$\pm$0.5 & 39.9$\pm$0.5 
        \\
        
        \multirow{-2}{*}{CAUM} 
        
        & \cellcolor{Gray} LF 
        & \cellcolor{Gray} 62.4$\pm$0.8 
        & \cellcolor{Gray} 63.5$\pm$0.8 
        
        & \cellcolor{Gray} 33.7$\pm$0.6 
        & \cellcolor{Gray} 33.7$\pm$0.7 
        
        & \cellcolor{Gray} 31.8$\pm$0.5 
        & \cellcolor{Gray} 31.8$\pm$0.8 
        
        & \cellcolor{Gray} 38.2$\pm$0.5 
        & \cellcolor{Gray} 38.0$\pm$0.7 

        & \cellcolor{Gray} 53.1$\pm$0.3 
        & \cellcolor{Gray} 65.9$\pm$0.2 
        
        & \cellcolor{Gray} 34.5$\pm$0.4 
        & \cellcolor{Gray} 34.5$\pm$0.1 
        
        & \cellcolor{Gray} 32.6$\pm$0.3 
        & \cellcolor{Gray} 32.8$\pm$0.1 
        
        & \cellcolor{Gray} 39.2$\pm$0.3 
        & \cellcolor{Gray} 39.3$\pm$0.1 
        \\
        \bottomrule  
        
    \end{tabular}%
    }
\end{table*}

Table \ref{tab:reults} shows the performance on MINDsmall and MINDlarge for both \texttt{C-AG} (NPA, NAML, NRMS, LSTUR, CenNewsRec, and MINS) and \texttt{C-AW} models (DKN, CAUM), under four different configurations of our comparative evaluation framework: (i) user modeling with \texttt{EF} vs. \texttt{LF}, combined with (2) training with \texttt{CE} vs. \texttt{SCL} objective. We next dissect the results along the three axes of our framework (\S\ref{sec:methodology}): user modeling, click behavior fusion, and  training objectives.   

\vspace{1.4mm} \noindent \textbf{Candidate-Agnostic vs. Candidate-Aware \texttt{NNR}s.} 
We analyze \texttt{C-AG} vs.\,\texttt{C-AW} models under their default \texttt{EF} configuration, since with \texttt{LF} all models become candidate-agnostic. CAUM, with the most complex and candidate-aware UE, generally outperforms all other models under both training regimes (\texttt{CE} and \texttt{SCL}) and for most evaluation metrics. The gaps are particularly prominent on the large training dataset, MINDlarge, w.r.t. the AUC metric. This result alone could mislead to a conclusion that more complex, candidate-aware user modeling is necessary for better recommendation. The fact that (1) DKN, as the other \texttt{C-AW} model in our evaluation -- generally performs much worse than \texttt{C-AG} models, as well as that (2) our \texttt{LF} models variants with trivial, parameterless UEs match or surpass the performance of CAUM with \texttt{EF}, undermine this conclusion.           
With the exception of DKN, all other models exhibit better performance when trained on the larger MINDlarge dataset.  
NAML and MINS are, however, competitive (except w.r.t. AUC metric) on MINDsmall, but fall behind CAUM on MINDlarge, suggesting that CAUM's elaborate UE benefits the most from more training data. 

One confounding factor that we do not control for, however, and which warrants a mindful comparison of the results, is that models differ not just in UE, but also in NE components, i.e., w.r.t. how they encode news and which features they use as input. 
For example, NAML and MINS, with an identical NE, achieve similar performance on MINDsmall. On MINDlarge, however, the more complex UE of MINS brings substantial gains over the simpler UE of NAML (but only under standard \texttt{EF} fusion and \texttt{CE} training).

\vspace{1.4mm} \noindent \textbf{Early vs. Late Click Behavior Fusion.}
Replacing complex \texttt{EF}-based UEs with the simple parameterless \texttt{LF} that we propose brings substantial performance gains across the board. Averaged across all models and both training objectives, \texttt{LF} brings massive gains of 5.58 and 4.63 MRR points on MINDsmall and MINDlarge, respectively. Equally importantly, with \texttt{LF} -- i.e., with the same parameterless UE -- models exhibit mutually much more similar performance than under \texttt{EF}, with other models generally closing the gap to CAUM. This suggests that \texttt{LF} makes differences in NE architectures across models less consequential, thus not only simplifying UE with parameterless averaging of clicked news embeddings, but also allowing for simpler news encoders.

\vspace{1.4mm} \noindent \textbf{Cross-Entropy vs. Supervised Contrastive Loss.} Overall, we find \texttt{SCL} to be a viable alternative to the common cross-entropy based classification with negative sampling (compare columns \texttt{CE} and \texttt{SCL} across evaluation metrics in Table \ref{tab:reults}).   
\texttt{SCL} brings large gains over \texttt{CE} in terms of AUC (+8.26 points on MINDsmall and +12.14 points on MINDlarge, averaged across all models, in both \texttt{EF} and \texttt{LF} variants). This suggests that, \texttt{SCL} leads to better separation of clicked and not clicked news in the representation space. In contrast, \texttt{SCL} falls slightly behind \texttt{CE} according to ranking measures, MRR and nDCG (-1.54 and -1.78 MRR points on MINDsmall and MINDlarge, respectively). We hypothesize that this is because of hard negatives -- news not clicked by the user that resemble user's clicked news -- for which \texttt{CE} more directly signals irrelevance: these likely become highly-ranked false positives for \texttt{SCL}-trained models.        

\begin{figure}[h]
  \centering
  \includegraphics[width=\linewidth]{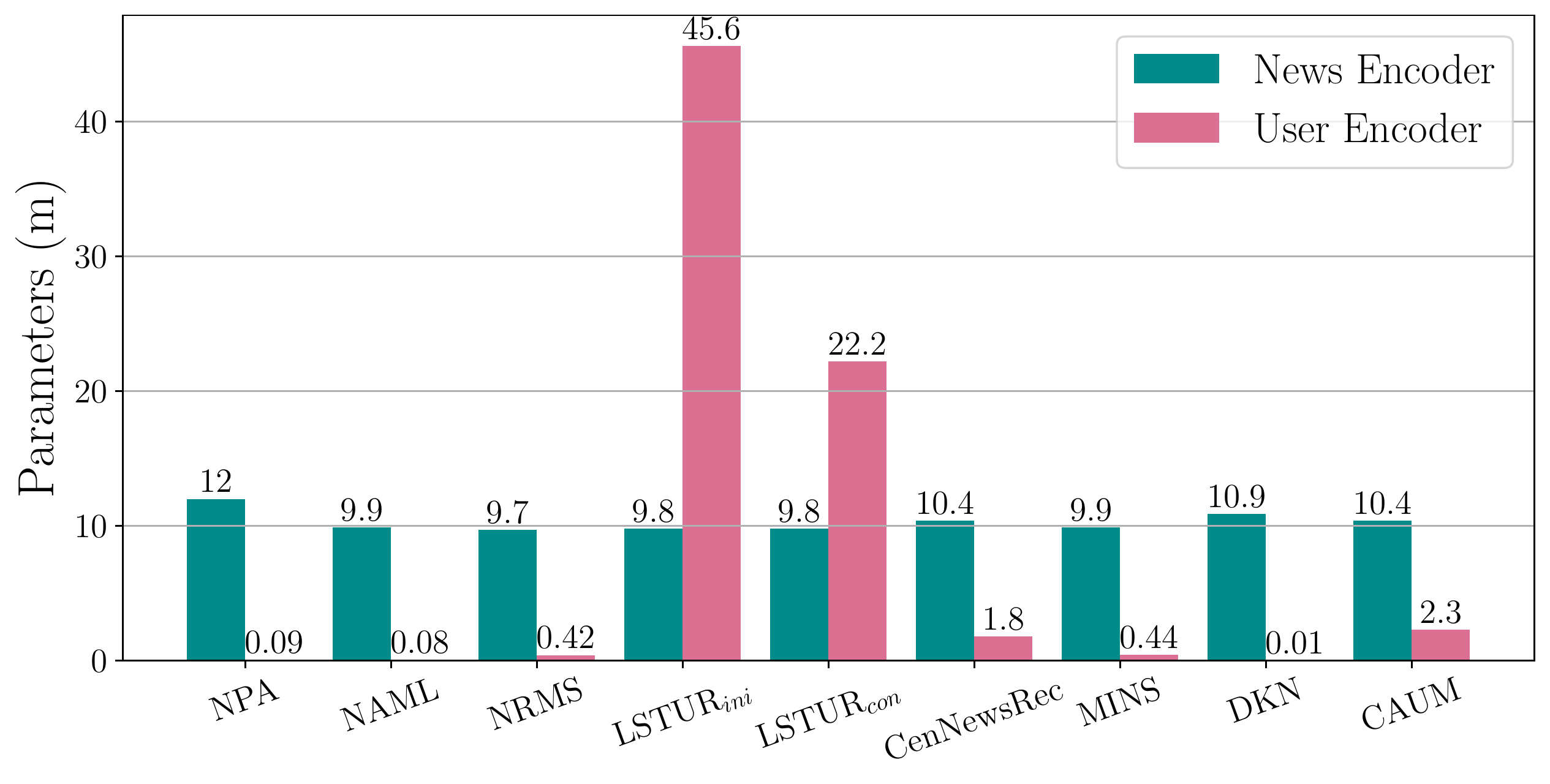}
  \caption{Number of model parameters (in millions).}
  \label{fig:parameters_ef}
  \vspace{-1em}
\end{figure}

\vspace{1.4mm} \noindent \textbf{Model Size.}
Finally, we quantify the reduction in model parameters that \texttt{LF} brings w.r.t. \texttt{EF}. Figure \ref{fig:parameters_ef} shows the number of trainable parameters in original \texttt{EF} configurations, on MINDsmall.\footnote{For some models, e.g., LSTUR with its user embedding matrix, the number of parameters depends on the size of the training data.} 
While the NE accounts for the majority of parameters in most models, the plot shows that the proportion of UE parameters is non-negligible for several models, and largest by a wide margin for LSTUR. With a parameterless UE, along with performance gains, \texttt{LF} brings a relative reduction of model size of 14.7\%, 18.1\%, and massive 82.3\% for CenNewsRec, CAUM, and LSTUR\textsubscript{ini}, respectively.

\section{Conclusion}
\label{sec:conclusion}

Rapid development of personalized neural news recommenders hinders fair comparative model evaluations and systematic analyses of design choices. 
In this work we introduce a unified framework for neural news recommendation focusing on three crucial design dimensions of \texttt{NNR}: (i) candidate-awareness in user modeling, (ii) click behavior fusion, and (iii) training objectives. Extensive evaluation of a wide range of recent state-of-the-art models reveals that \texttt{NNR} can be drastically simplified: replacing complex user encoders with parameterless aggregation of clicked news embeddings brings substantial performance gains across the board, reducing at the same time model complexity. Further, we show that contrastive learning can be a viable alternative to standard classification-based (cross-entropy) loss. We hope that our findings will inspire more transparent \texttt{NNR} evaluation, including systematic model ablations to uncover components that drive the performance.


\begin{acks}
The authors acknowledge support by the state of Baden-Württemberg through bwHPC. Andreea Iana was supported by the ReNewRS project grant, which is  funded by the Baden-Württemberg Stiftung in the Responsible Artificial Intelligence program.
\end{acks}

\bibliographystyle{ACM-Reference-Format}
\bibliography{references}


\end{document}